\documentclass[11pt]{article}
\usepackage{amsmath}
\usepackage{amssymb}
\newcommand{\be}{\begin{equation}}
\newcommand{\ee}{\end{equation}}
\newcommand{\beq}{\begin{equation}}
\newcommand{\eeq}{\end{equation}}
\newcommand{\ba}{\begin{eqnarray}}
\newcommand{\ea}{\end{eqnarray}}

\begin{document}
\baselineskip=15.5pt
\pagestyle{plain}
\setcounter{page}{1}



%

\def\ads{{\it AdS}}
\def\adsp{{\it AdS}$_{p+2}$}
\def\cft{{\it CFT}}

\newcommand{\ber}{\begin{eqnarray}}
\newcommand{\eer}{\end{eqnarray}}

\newcommand{\beqar}{\begin{eqnarray}}
\newcommand{\eeqar}{\end{eqnarray}}
\newcommand{\lm}{\lambda}\newcommand{\Lm}{\Lambda}
\newcommand{\ldm}{Lorentz/diffeomorphism}


\newcommand{\nonu}{\nonumber}
\newcommand{\oh}{\displaystyle{\tfrac{1}{2}}}
\newcommand{\dsl}
  {\kern.06em\hbox{\raise.15ex\hbox{$/$}\kern-.56em\hbox{$\partial$}}}
\newcommand{\sqg}{{\sqrt{-g}}}
\newcommand{\gumn}{{g^{\mu \nu}}}
\newcommand{\gdmn}{{g_{\mu \nu}}}
\newcommand{\Gdmn}{{G_{\mu \nu}}}
\newcommand{\goo}{{g^{00}}}
\newcommand{\gdoo}{{g_{00}}}
\newcommand{\eeqarr}{\end{eqnarray}}
\newcommand{\pa}{\partial}
\newcommand{\gab}{{g^{\alpha \beta}}}
\newcommand{\Roo}{{R_{00}}}
\newcommand{\Rrr}{{R_{rr}}}
\newcommand{\Rthet}{{R_{\theta \theta}}}
\newcommand{\nupp}{{\nu^{\prime\prime}}}
\newcommand{\nup}{{\nu^{\prime}}}
\newcommand{\labpp}{{\lambda^{\prime\prime}}}
\newcommand{\labp}{{\lambda^{\prime}}}
\newcommand{\Ttoo}{{\tilde{T}_{00}}}
\newcommand{\Too}{{T_{00}}}
\newcommand{\Tij}{{T_{ij}}}
\newcommand{\Ttij}{{\tilde{T}_{ij}}}
\newcommand{\Toi}{{T_{0i}}}
\newcommand{\Ttoi}{{\tilde{T}_{0i}}}
\newcommand{\Tmn}{{T_{\mu \nu}}}
\newcommand{\Ttmn}{{\tilde{T}_{\mu \nu}}}
\newcommand{\ZZ}{{\rm \kern 0.275em Z \kern -0.92em Z}\;}
\newcommand{\Rie}{{R^\al_{\ \beta \ \mu \nu}}}
\newcommand{\tilG}{{\tilde{G}}}
\newcommand{\tGbn}{{\tilde{G}_{\beta \nu}}}
\newcommand{\varmu}{{\varepsilon^{\mu \al \beta}}}
\newcommand{\varnu}{{\varepsilon^{\nu \al \beta}}}
\newcommand{\rtwo}{{^{(2)} R}}

\newcommand{\al}{\alpha}
\newcommand{\ga}{\gamma}
\newcommand{\Ga}{\Gamma}
\newcommand{\de}{\delta}
\newcommand{\De}{\Delta}
\newcommand{\ep}{\varepsilon}
\newcommand{\la}{\lambda}

\newcommand{\diff}{diffeomorphism~}
\newcommand{\cov}{covariant~}
\newcommand{\doe}{This work is  supported  by funds provided by the U.S Department of Energy (DOE)  under cooperative research agreements DE-FG02-05ER41360.}
\begin{titlepage}

\begin{center} \Large \bf Weyl Invariant Dynamics in 3 Dimensions\footnote{PASCOS 2005, Gyeongju, Korea, May-June 2005}

\end{center}

\vskip 0.3truein
\begin{center}
R. Jackiw

\vspace{0.3in}
Department of Physics\\
 Massachusetts Institute of
Technology \\
Cambridge, MA 02139, USA

\vspace{0.3in}

\end{center}
\vskip 1truein

\begin{center}
\bf ABSTRACT
\end{center}
We analyze a variety of Weyl invariant dynamical problems in three dimensions.

\vskip2.6truecm
\vspace{0.3in}

\smallskip

\vspace{0.3in}
\leftline{MIT-CPT-3681}
\smallskip
\end{titlepage}
\setcounter{footnote}{0}

\section{Introduction}
Weyl transformations and Weyl symmetries arise in discussions of field theories on curved space-time, whose non-trivial metric tensor $\gdmn$, changes under a Weyl transformation into a conformally related metric tensor.
\be
\gdmn \to \lambda^2 \, \gdmn
\label{eq:1}
\ee
Here $\la$ is an arbitrary space-time dependent function. Matter fields transform according to dimension-specific rules that also depend on the type of field. For example, a scalar  field $\varphi$ in $d$ dimensions, $d > 2$, transforms as 
\be
\varphi \to \la^{\frac{2-d}{2}}\, \varphi.
\label{eq:2}
\ee

Invariance under Weyl transformations is the curved space analog of flat space conformal invariance. It is easy to show that a diffeomorphism and Weyl invariant theory descends in flat space to a conformal invariant theory \cite{jack1}. Interest in Weyl invariance arises from the fact that the ``standard" particle physics model is Weyl invariant (in curved space time) save for its gauge symmetry breaking sector --- but little is certain about the ultimate features of gauge symmetry breaking.

However, Einstein theory is not Weyl invariant, because it is based on the Weyl non-invariant (contracted) Riemann tensor. The 4-index Riemann tensor $\Rie$ in dimension $d \geqslant 4$ can be decomposed into the (traceless) Weyl tensor $C^\mu_{\ \beta\ \mu \nu}$, which is Weyl invariant [it remains unchanged under the redefinition (\ref{eq:1}), and vanishes if and only if the space time is conformally flat], supplemented by Weyl-non-invariant contractions of the Riemann tensor:
\be
\Rie = C^\alpha_{\ \beta\ \mu \nu} + \frac{1}{d-2}\ \big(\de^\al_\mu\, \tGbn - \de^\al_\nu\, \tilde{G}_{\beta \mu} + g_{\beta \nu} \tilG^\al_\mu - g_{\beta \mu}\, \tilG^\al_\nu \big)
\label{eq:3}
\ee
where
\ba
\tilG_{\mu \nu} &\equiv& R_{\mu \nu} - \frac{1}{2(d-1)}\ \gdmn\, R \nonumber\\[.75ex]
R_{\mu \nu} &\equiv& R^\al_{\ \mu\ \al \nu} , \quad R \equiv \gumn\, R_{\mu \nu} \nonumber
\ea
Note that when the Einstein tensor $\Gdmn \equiv R_{\mu \nu} - \frac{1}{2}\, \gdmn\, R$ vanishes, so does $\tilG_{\mu \nu}$, but the Weyl tensor need not vanish, allowing for non-trivial vacuum geometries in general relativity, {\it e.g.} Schwarzschild space-time.

For $d \geqslant 4$, Weyl invariant gravity theory can be constructed, but it is very unwieldy owing to the absence of simple Weyl-invariant low-rank tensors or scalars (recall traces of the Weyl tensor vanish). However, new possibilities arise in   $d = 3$, where the Weyl tensor vanishes identically. The Riemann tensor is determined solely by $\gdmn$ and $\tilG_{\mu \nu}$, so that a vanishing Einstein tensor renders the space time to be flat.

While 3-dimensional space-time does not use a Weyl invariant tensor to describe its full (Riemann) curvature, another tensor replaces the absent Weyl tensor as a template for conformally flat space-times ({\it i.e.} it vanishes if and only if space-time is conformally flat). This is  the Cotton tensor, unique to three dimensions and given by a covariant curl of $\tilG^\nu_\beta \equiv R^\nu_\beta - \frac{1}{4}\ \delta^\nu_\beta \, R$.
\ba
C^{\mu \nu} &=& \frac{\varmu}{\sqrt{g}}\ D_\alpha\, \tilG^\nu_\beta \nonumber\\[.75ex]
&=& \frac{1}{2\sqrt{g}}\ \big(\varmu\, D _\alpha \, R^\nu_\beta + \varnu\, D_\alpha\, R^\mu_\beta\big) \\[.75ex]
&=& C^{\nu \mu} \nonumber
\label{eq:4}
\ea
The second equality, exhibiting the $\mu \leftrightarrow \nu$ symmetry, follows from the first by the Bianchi identity, which also ensures that $C^{\mu \nu}$ is covariantly conserved. Because $C^{\mu \nu}$ is symmetric,  covariantly conserved and traceless, one expects that it is the variational derivative with respect to $\gdmn$ (symmetry!) of a covariant scalar (covariant conservation!), which is also Weyl invariant (traceless!). This is indeed so, where the quantity to be varied is the gravitational Chern-Simons term.

To construct the gravitational Chern-Simons term, we recall first the gauge theoretic Chern-Simons term presented in terms of $A_\mu$, a Lie-algebra matrix valued gauge connection \cite{jack2}.
\be
CS(A) = \varepsilon^{\mu \nu \omega}\ t r\ \bigg(\frac{1}{2}\ A_\mu\, \partial_\nu \, A_\omega + \frac{1}{3}\ A_\mu \, A_\nu\, A_\omega\bigg)
\label{eq:5}
\ee
Next we call attention to an analogy between $(A_\mu)^\alpha_{\ \beta}$, where $(\al, \beta)$  are matrix indices, and $\Gamma^{\  \al}_{\mu \ \, \beta}$. The analogy rests on the fact that a diffeomorphism transformation $ x \to X (x)$ on $\Gamma^{\  \al}_{\mu \ \, \beta}$ formally can be viewed as a \diff on a covariant vector (index $\mu$) and a ``gauge transformation" in the  ``matrix indices" $(\al, \beta)$ with the ``gauge function" $U^{\ \al}_{\beta} = \frac{\partial X^\al}{\partial x^\beta}$. Moreover, the Riemann curvature $\Rie$ is constructed from $\Gamma^{\ \al}_{\mu \ \, \beta}$ in the same way as the gauge curvature $(F_{\mu \nu})^\alpha_{\ \beta}$ is constructed from $(A_\mu)^\alpha_{\ \beta}$. (The analogy can be extended to further correspondences between gauge theoretic and geometrical entities \cite{jack3}.)

With the above analogy, we can immediately construct the  gravitational Chern-Simons term \cite{jack2}.
\be
CS(\Gamma) = \varepsilon^{\mu \nu \omega}\ \bigg(\frac{1}{2}\ \Ga^{\ \al}_{\mu\ \, \beta}\, \partial_\nu\, \Ga^{\ \beta}_{\omega\ \, \al} + \frac{1}{3}\ \Ga^{\ \alpha}_{\mu\ \, \beta}\, \Ga^{\ \beta}_{\nu\ \, \ga} \, \Ga^{\ \ga}_{\omega \ \, \al}\bigg)
\label{eq:6}
\ee
It follows that
\be
\delta \int d^3 x\, CS(\Ga) = \frac{1}{2} \ \int d^3 x\ \sqrt{g}\ C^{\mu \nu}\, \delta \, \gdmn.
\label{eq:7}
\ee

Symmetry and conservation of $C^{\mu \nu}$ indicate that it can be added to the usual  Einstein term for an extended  3-dimensional gravity theory.  The extended vacuum equation is 
\be
\Gdmn + \frac{1}{m} \ C_{\mu \nu} = 0,
\label{eq:8}
\ee
where $1/m$ is the strength of the extension. For dimensional balance, $m$ must carry dimensionality of mass.

The extended theory, unlike the Einstein theory, possesses propagating degrees of freedom carrying mass $m$. The limit $m\to \infty$ returns (\ref{eq:8}) to the Einstein model, whose absent excitations are seen to decouple from the extended theory since their mass becomes infinitely large.

Here we shall consider the opposite limit $m\to 0$, where only the Cotton tensor survives, giving rise to a Weyl invariant gravity theory in 3-dimensional space-time, some of whose solutions we shall now explore.

\section{Sourceless Solutions}
The sourceless, Weyl invariant equation
\be
C^{\mu \nu} = 0
\label{eq:9}
\ee
at first appears without structure, since all solutions are coordinate transformations of arbitrary conformally flat space times.
\be
\gdmn = \rho \, \eta_{\mu \nu}
\label{eq:10}
\ee
However, structure emerges if we also demand the metric, as a function of $(t, x, y)$ in a Kaluza-Klein parameterization, be independent of one coordinate, which we take to be the second spatial one, $y$ \cite{jack4}.

The emergent equations are presented by first parameterizing the 3-dimensional metric tensor, in the Kaluza-Klein fashion, as 
\be
\mbox{3-d metric tensor} = \chi \ \left(\begin{array}{cl}g_{a b} - A_a A_b & -A_a \\ -A_b & -1\end{array}\right).
\label{eq:11}
\ee
Here $g_{ab}$ is the 2-dimensional metric tensor for the $(t,x)$ space-time, $A_a$ a 2-component 2-vector, and $\chi$ is a scalar. Also, we demand that all quantities be $y$-independent. It follows that under 3-dimensional diffeomorphisms, which preserve that requirement, $g_{ab}, A_a \ \mbox{and}~\chi$ transform properly as a 2-dimensional tensor, vector and scalar, respectively,  and $A_a$ undergoes an Abelian gauge transformation. The action for 2-dimensional motion involves the Chern-Simons term, integrated over the remaining 2-dimensional space-time.
\be
\int d^2 x\, CS(\Ga) = - \frac{1}{2}\ \int d^2x\ \sqrt{-g}\ \bigg({^{(2)} R}\,  F + F^3\bigg)
\label{eq:12} 
\ee
Here ${^{(2)} R}$ is the 2-dimensional scalar curvature and $F$ is the (dual) ``field" associated with the ``potential"  $A_a$.
\be
F_{ab} \equiv \partial_a\, A_b - \partial_b\, A_a = \varepsilon_{ab}\ \sqrt{-g}\ F
\label{eq:13}
\ee
where $\varepsilon_{ab}$ is anti-symmetric and $\varepsilon_{01} = -1$.

The disappearance of $\chi$ is due to the Weyl symmetry of the Chern-Simons action. Alternate presentations of (\ref{eq:12}) highlight its topological nature,
\begin{subequations}\label{eq:14}
\be
\int d^2 x \, CS(\Ga) \propto \int d A \ \bigg(\rtwo + F^2\bigg) \ , \ A\equiv A_a\, d x^a,
\label{eq14a}
\ee
and its axion-like structure
\be
\int d^2 x \, CS(\Ga) \propto \int d^2 x\, \Theta\, \varepsilon^{ab}\, F_{ab}, \quad \Theta = \rtwo + F^2.
\label{eq:14b}
\ee
\end{subequations}
The equations of motion that follow from (\ref{eq:9}) or from (\ref{eq:12}) read
\begin{subequations}\label{eq:15}
\begin{alignat}{2}
&\rtwo+ 3 F^2 = \mbox{constant} \equiv C\label{eq:15a}\\
&0 = D^2 \, F - C\, F + F^3\label{eq:15b}\\
&0= (D_a D_b - \frac{1}{2}\ g_{ab}\, D^2)\ F .
\label{eq:15c}
\end{alignat}
\end{subequations}
Eq. (\ref{eq:15a}) is gotten by varying $A_a$ in (\ref{eq:12}) and integrating once; (\ref{eq:15b}) and (\ref{eq:15c}) result when the variation of $g_{ab}$ is presented in terms of its trace and traceless parts. Note that while (\ref{eq:12}) has the appearance of a ``dilaton" gravity with $F$ as the dilaton, the crucial difference is that for us $F$ is not a fundamental variable; rather $A_a$ is the independent variable.

We observe that the equations enjoy the ``symmetry" $F \leftrightarrow -F$, but the action is not invariant --- it changes sign. Solutions can be classified according to their response to this symmetry. First there are the homogenous solutions, which either preserve the symmetry
\be
\mbox{``symmetric" solution:} \qquad F=0, \rtwo = C, 
\label{eq:16}
\ee
or break it (if $C> 0$)
\be
\mbox{``symmetry breaking" solutions:} \quad F = \pm \sqrt{C}, \ \rtwo = -2 C < 0.
\label{eq:17}
\ee
The two correspond to a 2-dimensional deSitter or anti-deSitter space-time, respectively.

Additionally there exists a ``kink" solution, which interlopes between the two symmetry breaking solutions, $F = \pm \sqrt{C} $.
\begin{subequations}\label{eq:18}
\begin{alignat}{2}
\mbox{``kink" solution:}\qquad &F = \sqrt{C} \tanh\ \frac{\sqrt{C}}{2}\ x \label{eq:18a}\\
& \rtwo  = -2C + \frac{3C}{\cosh^2\frac{\sqrt{C}}{2} \ x}
\label{eq:18b}
\end{alignat}
\end{subequations}
[There also exist more general solutions, which depend on two integration constants: (i) origin of the $x$ coordinate (here set to zero), (ii) selection of a specific first integral (here corresponding to the kink). The local geometry of these more general solutions is the same as that of the kink \cite{jack5}. We omit from discussion the trivial case $C=0$.]

We may take a 3-dimensional viewpoint towards these geometries. The 3-dimensional line element for the ``symmetric" solutions (\ref{eq:16}) is 
\begin{subequations}\label{eq:19}
\ba
C>0: ds^2 &=& \frac{2}{C}\ \Bigg[\Bigg(\frac{dt}{t}\Bigg)^2 - \Bigg(\frac{dx}{t}\Bigg)^2\Bigg] - d y^2, \label{eq:19a}\\[1ex]
C<0: ds^2 &=& \frac{2}{|C|}\ \Bigg[\Bigg(\frac{dt}{x}\Bigg)^2 - \Bigg(\frac{dx}{x}\Bigg)^2\Bigg] - d y^2,
\label{eq:19b}
\ea
\end{subequations}
for the symmetry breaking solution (\ref{eq:17})
\be
C>0\qquad  ds^2 = - \frac{2}{\sqrt{C}\ x}\ dt dy - \bigg(\frac{dx}{\sqrt{C}\ x}\bigg)^2 - dy^2,
\label{eq:20}
\ee
for the kink solution (\ref{eq:18})
\be
C>0\qquad  ds^2 = -\frac{2}{\cosh^2\, \frac{\sqrt{C}\ x}{2}}\ dt dy - (dx)^2 - (dy)^2.
\label{eq:21}
\ee

For all of these $C^{\mu \nu}$ vanishes, so that the 3-dimensional space-time is conformally flat. Therefore with a change of coordinates, it must be possible to present these line elements as $\rho\, d X^\mu\ dX^\nu\ \eta_{\mu \nu}$. Also the 3-dimensional scalar curvature $R$ (with $\chi=1$) is given by
\[
R = \rtwo + \frac{1}{2} \ F^2.
\]
The needed coordinate transformation can be found, with the results for the symmetric solutions
\ba
\hspace{2in} \rho &=& \frac{2}{C(T^2 - Y^2)} , \quad R = C > 0, \hspace{1.88in} (19'a) \nonumber\\
\hspace{2in} \rho &=& \frac{2}{|C| (X^2 + Y^2)} ,\quad R = C< 0. \hspace{1.80in} (19'b) \nonumber
\ea
These space-times possess four 3-dimensional Killing vectors, spanning $ SO(2,1) \times SO(2)$ and describe 2-dimensional deSitter $(C>0)$ or anti-deSitter $(C<0)$ space embedded in three dimensions. For the symmetric breaking solution we have
\be
\hspace{2in} \rho = \frac{4}{CX^2} , \quad R = -\frac{3}{2}\ C<0, \hspace{2.25in} (20') \nonumber
\ee
with six 3-dimensional Killing vectors spanning $SO(2,1) \times SO(2,1)$. This is the maximally symmetric $(R^\mu_\nu = \frac{1}{3}\ \delta^\mu_\nu\, R = -\frac{1}{2}\ \delta^\mu_\nu\, C)$ 3-dimensional anti-deSitter space. Finally for the kink we have
\be
\hspace{.46in} \rho = \frac{4}{1-C(T-Y)^2 + CX^2} ,\quad R =  -\frac{3C}{2} + \frac{5C}{2\cosh^2\frac{\sqrt{C}x}{2}}. \hspace{2in} (21') \nonumber
\ee

Note the similarity between the present curved space kink and the kink that solves the flat space version of (\ref{eq:15b}).
\ba
\Box f - C f + f^3 =0 \nonumber\\
f = \sqrt{C} \ \tanh\ \sqrt{\frac{C}{2}} \ x
\label{eq:22}
\ea
The only difference between this and (\ref{eq:18a}) is  the scale of $x$.\!\! This is an instance of a general phenomenon: if a $f(x)$ is  a static kink solution to 
\begin{alignat}{2}
&\Box f + V^\prime (f) = - f^{\prime \prime} + V^\prime (f) = 0 \nonumber\\
&\Rightarrow \ f^\prime = \sqrt{2 V (f)}
\label{eq:23}
\end{alignat}
then the 2-dimensional equations
\ba
D^2 \, F + V^\prime (F) = 0,\nonumber\\
(D_a D_b - \frac{1}{2}\ g_{ab}\, D^2)\ F =0,
\label{eq:24}
\ea
are solved by 
\be
F(x) = f (x/\sqrt{2}),
\label{eq:25}
\ee
and
\ba
ds^2 &=& V(F)\, dt^2 - dx^2, \nonumber\\
\rtwo &=& - V^{\prime \prime} (F).
\label{eq:26}
\ea

These results have been extended by coupling fermions to the system, either with or minimally without supersymmetry. As expected the Dirac equation possesses zero modes in the field of the kink \cite{jack6}. 

The flat space space kink has found physical application as a defect in the binding pattern of 1-dimensional polymers, like polyacetylene. Moreover fermions interacting with these defects undergo charge fractionalization, as a consequence of the zero modes \cite{jack7}. The physical correlatives of the curved space kink remain as yet unknown.

\section{Solutions with Sources}
Another dynamical problem that we have analyzed involves an scalar field energy-momentum tensor $T_{\mu \nu}$ as a source \cite{jack8}.
\be
C_{\mu \nu} = k \ T_{\mu \nu}
\label{eq:27}
\ee
$T_{\mu \nu}$ is ``improved" so that it is traceless, just as is $C_{\mu \nu}$. Thus we take
\ba
T_{\mu \nu} &=& \partial_\mu \, \varphi\, \partial_\nu\, \varphi - \frac{1}{2} \ \gdmn\, g^{\al \beta} \, \partial_\al\, \varphi\, \partial_\beta \varphi \nonumber\\
&& + \frac{1}{8} (g_{\mu \nu} D^2 - D_\mu D_\nu + G_{\mu \nu})\, \varphi^2,
\label{eq:28}
\ea
where the last term is the ``improvement" term, which survives in the flat-space limit. This energy momentum tensor arises when the scalar field is non-minimally, but conformally coupled to the geometry.
\be
\mathcal{L} = \sqrt{g}\ \bigg(\frac{1}{2}\ \gumn \, \partial_\mu \, \varphi\, \partial_\nu\, \varphi + \frac{1}{16}\ R\, \varphi^2\bigg)
\label{eq:29}
\ee

The dynamics in (\ref{eq:27}) and (\ref{eq:28}) supports ``pp waves", {\it viz}.  the line element
\begin{alignat}{2}
& ds^2 = F(u,y) \ d\, u^2 + 2\, du\,  dv - dy^2\\
& (u, v) = \frac{1}{\sqrt{2}} \ (t \pm x) \nonumber
\label{eq:30}
\end{alignat}
with
\ba
F(u, v) &=& f(u)\, exp \, \bigg[\frac{\kappa y}{8\sigma (u)}\bigg] - \frac{\ddot{\sigma} (u)}{\sigma (u)}\ y^2\nonumber\\
&& \varphi = 1/\sqrt{\sigma(u)}.
\label{eq:31}
\ea
The Ricci  (equivalent in $d = 3$ to the full) curvature has only one non-vanishing component.
\be
R_{uu} = \frac{\kappa^2}{128\, \sigma^2(u)} \ f (u)\, exp\ \bigg[\frac{\kappa y}{8\sigma  (u)}\bigg] - \frac{\ddot{\sigma} (u)}{\sigma (u)}
\label{eq:32}
\ee
This solution may be generalized by adding a Weyl invariant self interaction to $\mathcal{L}: \mathcal{L}_I \sim \varphi^6$. Also non-minimal couplings as in (\ref{eq:29}), but with strength other than the conformal value of $1/16$, so that the energy-momentum tensor is not traceless, can still be consistently solved. Of course on the solution $T^\mu_{\ \mu}$ vanishes \cite{jack9}.

\doe

\end{document}